\documentclass[10pt]{article}   
\usepackage{osajnl2} 
\newcommand{\mlaser}{emitting laser}
\begin{document}

\title{Lasers for coherent optical satellite links with large dynamics}


\author{Nicola Chiodo, Khelifa Djerroud, Ouali Acef, Andr\'{e} Clairon, and Peter Wolf$^{*}$}
\address{LNE-SYRTE, Observatoire de Paris, CNRS, LNE, UPMC, 61 av. de l'Observatoire, \\ 75014 Paris, France}
\address{$^*$Corresponding author: Peter.Wolf@obspm.fr}

\begin{abstract}
We present the experimental realization of a laser system for
ground to satellite optical Doppler ranging at the atmospheric
turbulence limit. Such a system needs to display good frequency
stability (a few parts in $10^{-14}$) whilst allowing large and
well controlled frequency sweeps of $\pm$12 GHz at rates exceeding
100 MHz/s. Furthermore it needs to be sufficiently compact and
robust for transportation to different astronomical observation
sites where it is to be interfaced with satellite ranging
telescopes. We demonstrate that our system fulfills those
requirements and should therefore allow operation of ground to low
Earth orbit satellite coherent optical links limited only by
atmospheric turbulence.
\end{abstract}

\ocis{(140.3425)   Laser stabilization; (140.3600)  Lasers, tunable;
(120.3940)   Metrology; (120.3930) Metrological instrumentation;
(060.2605)   Free-space optical communication; (010.1330)   Atmospheric turbulence
}

\maketitle 


\section{Introduction} \label{Intro}

An increasing number of applications require laser sources with
low frequency noise, large frequency dynamics, or both. Examples
are LIDAR, optical communications to satellites, optical Doppler
ranging and telemetry to satellites or between satellites, optical
spectrum analyzers, coherent manipulation of atoms for quantum
processing, and low noise interferometric sensors.

In this work we consider one particular application: ground to
satellite optical Doppler ranging at the atmospheric turbulence
limit. That technique uses phase coherent tracking of an optical
signal reflected from a corner cube onboard a low Earth orbit
satellite, thereby measuring the optical distance variations
between the ground telescope and the satellite with a potential
for several orders of magnitude improvement over standard
satellite laser ranging techniques. More importantly, such
measurements, when limited by atmospheric turbulence, provide
valuable information on the phase and amplitude noise of that
effect (up and down link) of particular importance for phase
coherent optical communications \cite{Gregory} and long distance
optical free space clock comparisons \cite{Djerroud,Giorgetta}.
The latter is of particular interest in the context of next
generation atomic clocks for space applications in fundamental
physics, navigation, and geodesy.

In 2009 coherent optical Doppler ranging to a stationary ground
target has demonstrated the feasibility of the method and its
limitations from atmospheric turbulence on a 5 km horizontal link
\cite{Djerroud}. Furthermore, that experiment has allowed the
estimation of the residual noise from atmospheric turbulence in
long distance clock comparisons, with very promising results.
Those estimations have been confirmed experimentally in another 2
km horizontal ground link very recently \cite{Giorgetta}. However
similar experiments for ground-space links are still missing and
direly needed in order to demonstrate the feasibility under much
more difficult conditions (large Doppler dynamics, low return
power, high frequency stability requirements of the source). Of
interest in their own right for precision satellite ranging they
are also crucial for the characterization of atmospheric
turbulence phase and amplitude noise.

Here we report on progress towards the realization of that goal.
We describe the experimental development of the required lasers
that need to display state of the art frequency stability, whilst
being capable of a controlled frequency sweep spanning $> 20$ GHz
at a maximum rate exceeding 100 MHz/s in order to compensate the
Doppler effect due to satellite motion. We derive the requirements
on the laser stability from the expected turbulence noise and show
that our laser system meets those requirements.

The application described above requires high stability and
controlled and broad frequency tunability at the same time, which
is a challenge to many laser stabilization schemes (eg.
Fabry-Perot cavities or locking to atomic/molecular transitions).
A solution is to use large arm length unbalance fibre
interferometers \cite{Kefelian}. It has been demonstrated that
such systems display state of the art stability with simultaneous
sweep linearity \cite{Jiang}. However, the achieved sweep rates
(up to 40 MHz/s) and range ($< 1$ GHz) are insufficient for our
purposes. A system with much faster and broader tuning was
reported in \cite{Roos,Barber}, however with insufficient
stability for our purposes.

Our own system is comprised of two lasers each stabilized on an
unequal arm fibre interferometer closely inspired by \cite{Jiang}.
The system is less stable than the one reported in \cite{Jiang}
but still sufficient for our purposes, whilst having broader and
faster tuning capability. It is a compromise between compactness
and performance as it has to be moved frequently to astronomical
sites where it will be used to realize first coherent
ground-satellite optical Doppler ranging. It has recently been
transported from Paris to the Observatoire de la c\^{o}te d'Azur
in southern France where first preliminary tests with the 1.5 m
lunar/satellite ranging telescope have taken place. Further tests
are planned in 2014.

The paper is organized as follows. We first describe the
application ie. the planned experiment for a coherent optical
ground to satellite link, and derive the resulting requirements on
the lasers. We then provide the details of our laser system
designed and built for that purpose. This is followed by the
experimental results of the laser frequency stability in different
configurations (no sweep, linear sweep, non-linear sweep). Finally
we discuss those results and conclude.

\section{Requirements for the lasers}
\label{Req}

The laser system described here was developed within the scope of
the Mini-DOLL (Miniature Deep-space Optical Laser Link) project
that aims at demonstrating the feasibility and studying the phase
noise limitations of coherent ground to satellite optical links
across large distances as proposed for distant clock comparisons
and communication across the solar system \cite{SAGAS,OSS}.
Mini-DOLL uses a two way link with the light reflected off a
corner cube onboard low Earth orbit satellites (around 1000 km
altitude). Thus the signal crosses the atmosphere twice and is
subject to the large Doppler dynamics of low Earth orbit
satellites. Concerning the received power, we expect less than 1
pW power in the return signal due mainly to the two-way
configuration and the small size of the corner cubes ($\approx 3$
cm). This is only about an order of magnitude more than the
received power from 1 W emission from a spacecraft at 3
astronomical units from the Earth \cite{OSS} ie. if successful
Mini-DOLL demonstrates the feasibility of such a deep space link
with similar received power levels but significantly more adverse
Doppler conditions and a double (rather than single) crossing of
the atmosphere.

\begin{figure}[!htbp]
  \includegraphics[width=0.5\textwidth]{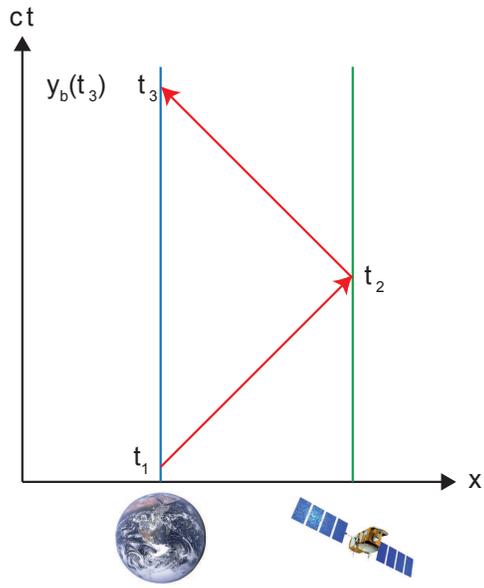}\\
  \caption{Space-time diagram of the measurement principle.}\label{fig_teory_satellite}
\end{figure}

Consider a ground to satellite link as shown in
fig.~\ref{fig_teory_satellite}. At the instant $t_{1}$ the laser
is emitted towards the satellite where, at time $t_{2}$, it is
reflected by the corner cube and received back at the ground
telescope at instant $t_{3}$. The measurement is the beatnote
(fractional frequency difference $y_b(t_3)$) between the local
laser and the signal returning from the satellite.

The measured beatnote can be written as
\begin{equation}\label{eq_1}
    y_b(t_{3})=y_{l}(t_{3})-y_{l}(t_{1})+2\frac{v_{s}}{c}(t_{2})-\frac{v_{g}}{c}(t_{1})-\frac{v_{g}}{c}(t_{3})+y_{atm}(t_{1})+y_{atm}(t_3)
\end{equation}
where $y_l(t)$ is the laser frequency at instant $t$ (normalized
to the nominal laser frequency), $v_s(t)$ and $v_g(t)$ are the
satellite and ground station velocities respectively projected
onto the line of sight, $y_{atm}(t)$ is the perturbation in
fractional frequency due to the crossing of the atmosphere
(turbulence), and $c$ is the velocity of light.

The aim is to determine the satellite motion (velocity terms in
(\ref{eq_1})) limited only by atmospheric turbulence ($y_{atm}$
terms in (\ref{eq_1})). For the laser stability this implies

\begin{equation}\label{eq_2}
    y_{l}(t_{3})-y_{l}(t_{1}) \leq y_{atm}(t_{1})+y_{atm}(t_3)
\end{equation}

which in terms of power spectral densities (PSD)
becomes\footnote{The PSD of $x(t)=n(t) \pm n(t-T)$, where $n(t)$
is a stochastic process, is given by $S_x(f) = \left( 1 \pm \cos
(2\pi f T) \right) S_n(f)$}

\begin{equation}\label{eq_3}
    \left( 1-\cos (2\pi f T_{13}) \right) S_{laser}(f) \leq \left( 1+\cos (2\pi f T_{13}) \right) S_{atm}(f)
\end{equation}

where $T_{13} = t_3-t_1$, and $S_{laser}(f)$, $S_{atm}(f)$ are the
noise PSD of the laser and the atmospheric perturbation
respectively.

Equation (\ref{eq_3}) defines our requirements on laser stability,
from an assumption on the noise induced by atmospheric turbulence
and from a knowledge of the ground to satellite distance ie.
$T_{13}$. Throughout the paper we will assume that the atmospheric
noise PSD is of the order of the one measured in
\cite{Djerroud,Giorgetta}, as justified in \cite{Djerroud}, and
that the satellites that we aim at are in low Earth orbit
typically at altitudes around 1000 km giving $T_{13} \approx 6.6$
ms.

Other requirements for our experiment come from the expected low
signal power and the large Doppler effect induced by satellite
motion. We expect that the signal power will be attenuated by at
least a factor 10$^{-13}$, due to divergence of the beam, small
size of the corner cube, attenuation of the atmosphere, losses in
the telescope and detection optics, and other effects. Emitting at
$\approx$ 30 W this implies a return power of the order of 1 pW.
To achieve a reasonable, shot noise limited, signal to noise ratio
the signal needs to be filtered in a narrow band of about 1 kHz.
The expected S/N ratio is then of the order 30 dB at our 1064 nm
wavelength, and expected to fluctuate due to atmospheric
scintillation. However, a typical satellite at 1000 km altitude
will induce Doppler shifts varying from +12 GHz to -12 GHz during
the satellite pass at a maximum rate of about 120 MHz/s. It is not
possible to measure a beatnote in a 1 kHz band that varies between
0 and 12 GHz at that rate, so we use the following approach to
overcome the Doppler effect. Our experiment is composed of 2
lasers, a local oscillator and an \mlaser{}. The \mlaser{} will be
at constant frequency while the laser frequency of the local
oscillator will be swept to compensate the Doppler shift of the
returning signal. We will sweep the frequency of the local
oscillator according to the satellite orbit prediction. The result
is a beatnote that varies sufficiently slowly to be tracked with a
1 kHz bandwidth (ie. it stays within a 1 kHz band for at least a
few ms).

In conclusion, the requirements on our laser system can be
summarized as follows:
\begin{itemize}
    \item Stabilize the \mlaser \ and local oscillator laser to a
    level satisfying (\ref{eq_3}).
    \item Allow for a controlled sweep of the local oscillator
    laser to compensate for the predicted (theoretical) Doppler
    shift due to satellite motion (+12 GHz to -12 GHz at a maximum
    rate of 120 MHz/s).
    \item Ensure that the difference between the actual laser
    frequency during the sweep and the predicted sweep has a
    stability satisfying (\ref{eq_3}).
    \item Allow for large emission power ($>$~10 W) without
    deteriorating the frequency stability.
\end{itemize}

\section{Laser setup}
\label{set-up}
\begin{figure}[!htbp]
  \includegraphics[width=\textwidth]{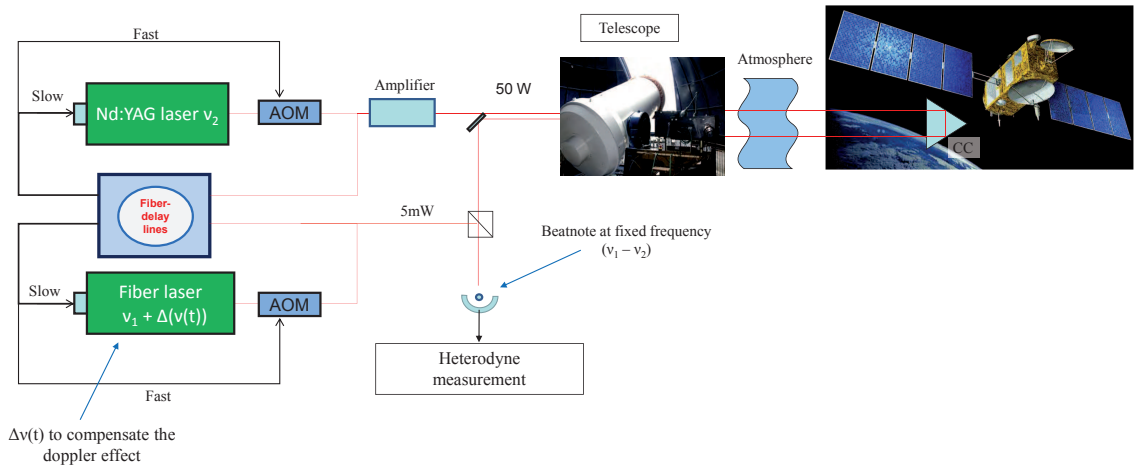}\\
  \caption{Laser set-up. The two lasers are stabilized on fibre delay lines. That stabilization is also used to vary the frequency of the local oscillator in order to compensate for the Doppler effect such that the beatnote is at constant frequency.}\label{figura_grande}
\end{figure}

In our setup we use two lasers one for emission (Nd:YAG laser +
Er:Yb multi-stage fibre amplifier) and a broadly tunable Fibre
Bragg Grating (FBG) laser as the local oscillator. The set-up is
shown schematically in figure \ref{figura_grande}.

\begin{figure}
  \includegraphics[width=\textwidth]{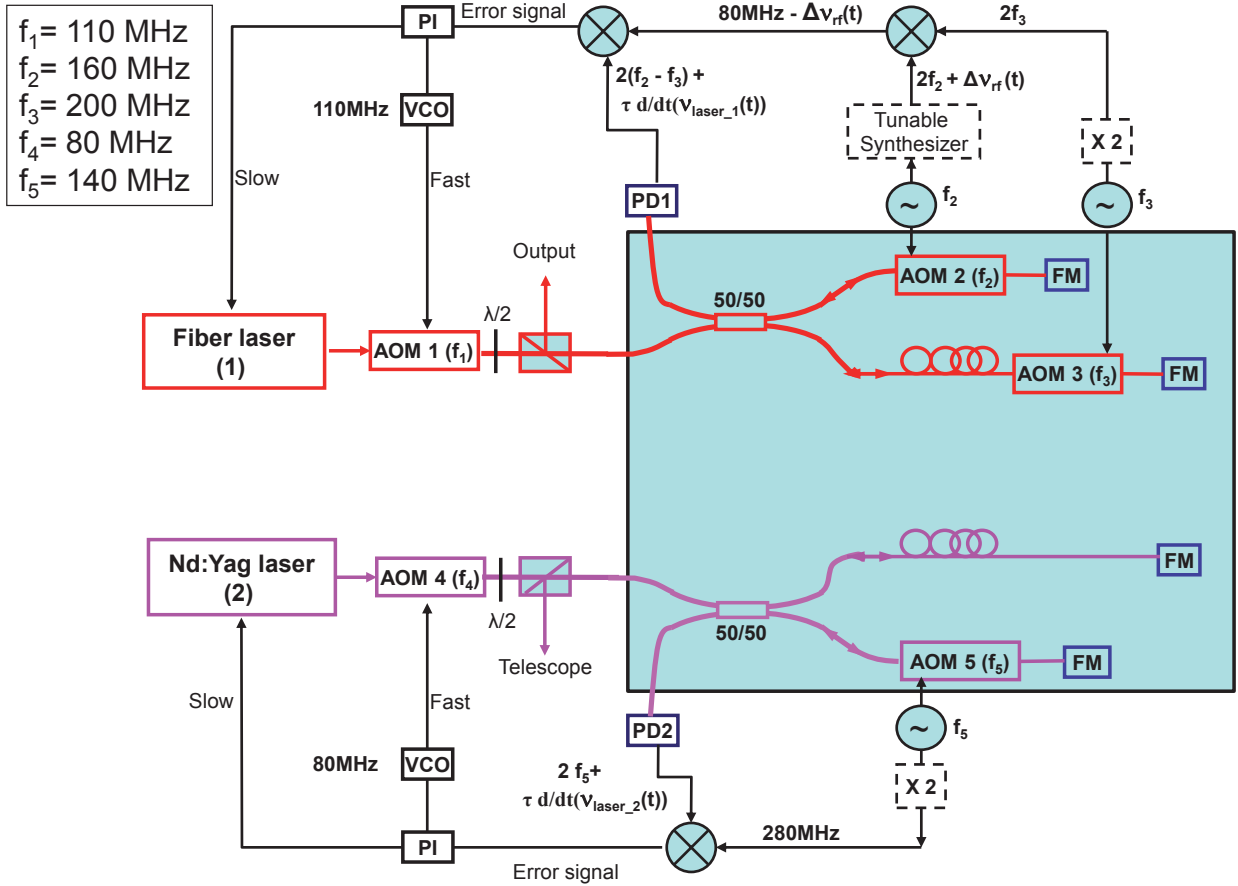}\\
  \caption{Laser stabilization scheme. AOM: Acousto-Optic Modulator; FM: Faraday Mirror; PD: Photodiode; VCO: Voltage Controlled Oscillator; PI: Proportional Integrator; $\tau$ = time delay due to arm inbalance. In closed loop operation, the error signals are zero, implying $\dot{\nu}_{laser}=0$ for the emission laser (YAG laser (2)) and $\dot{\nu}_{laser}(t)=\Delta \nu _{rf}(t)/\tau$ for the local oscillator (Fiber laser
  (1)).}\label{fig_rf}
\end{figure}

The stabilization and sweep control are based on the method
described in \cite{Jiang}, and the reader is referred to that
article for details. In our case we stabilize our two lasers on a
1 km (local oscillator) and 2 km (\mlaser) single mode fibre delay
line, which are both housed in the same temperature controlled and
seismically isolated box. The detailed scheme is shown in figure
\ref{fig_rf}.

Due to the arm imbalance $\tau$ ($\tau \approx 10 \mu$s and 20
$\mu$s for the 1 km and 2 km spools respectively) of the
Michelson-Morley interferometers, the beatnote frequency on
photodiodes PD1 and PD2 depends on $\tau \dot{\nu}_{laser}(t)$ ie.
on the laser frequency derivative. Therefore, by adding a
frequency $\Delta \nu_{rf}(t)$ to the demodulation signal of the
fibre laser, in closed loop operation (error signal = 0), its
frequency is varied (swept) according to the condition

\begin{equation}\label{eq_302}
    \dot{\nu}_{laser}(t)=\frac{\Delta \nu _{rf}(t)}{\tau}.
\end{equation}

The programmed rf frequency for the sweep $\Delta\nu_{rf}$ is
provided by a direct digital synthesizer (DDS), to which we upload
the predicted satellite Doppler shifts before the pass. During the
satellite pass (about 20 min) the required laser frequency
derivative $\dot{\nu}_{laser}$ varies from 0 to 120 MHz/s and back
which requires generating an appropriate ramp of $\Delta\nu_{rf}$
varying from 0 to 1.2 kHz according to equ. (\ref{eq_302}). The
DDS generates a new frequency every millisecond according to a
previously uploaded file. Additionally it allows for manual
frequency steps or time steps during the pre-programmed sequence
in order to be able to correct "in real time" for orbit prediction
errors. Any manual adjustments (their value and the time at which
they are applied) are logged in a file for later traceability.


The \mlaser{} is a Nd:YAG oscillator emitting at 1064 nm
(manufactured by Innolight). The maximum output power is 500 mW,
the spectral linewidth (over 100 ms) is 1 kHz. The wavelength can
be tuned by acting on the temperature by 30 GHz, and by acting on
the piezo-electric actuator (PZT), mounted on one mirror of the
laser cavity, by $\pm$100 MHz. The fibre amplifier is manufactured
by Nufern and provides 50 W maximum output. The local oscillator
is a DFB Yb:fiber laser emitting at 1064 nm (manufactured by NKT
photonics). The maximum output power is 100 mW, the spectral
linewidth inferior to 10 kHz. The wavelength can be tuned by
acting on the temperature by $\sim$120 GHz, and by acting on the
PZT by $\sim$106 GHz (wide piezo tuning option of NKT-photonics).
To sweep the laser for Doppler compensation, we have chosen to act
on the PZT of the local oscillator only. The typical optical power
sent into each interferometer is $\approx 50$ $\mu$W.

Different devices and techniques are used to ensure the thermal,
vibration and pressure isolation of the fiber spools. All these
effects worsen the performances of the stabilization system in
different ways. For thermal and pressure isolation we use 4 boxes
one inside the other. The 2 fiber spools are inserted in a
metallic box with diameter 30 cm and height 20 cm. This box is
inserted in another metallic box with cylindrical shape.

Inside this second box, but not in thermal contact with the box of
the fibers, are placed the 3 AOMs (Acousto Optic Modulators) of
the interferometers (AOM 2, 3, and 5 in fig. \ref{fig_rf}). The
second box is actively thermally stabilized by 4 peltier modules
attached on the external lateral surface. On the lid of the second
box are placed the 2 photodiodes.

The 2 metallic boxes are air-sealed to attenuate pressure
variations. The second box is placed inside another metallic box
of cubic shape. The inside of the cubic box is covered with 5 cm
thin acoustic absorber (polyurethane agglomerate) to attenuate
acoustic vibrations. The cubic box is mounted on an active
vibration table that filters the seismic vibrations between 10 and
100~Hz.

To dissipate the heat of the peltier elements, they are in thermal
contact with the ground by means of 4 twisted copper wires.
Finally the cubic box is housed inside a wooden cubic box to
reduce fluctuations of room temperature, to further attenuate
acoustic noise and to protect the complete set-up. The inside of
the wooden box is covered with 5 cm thick acoustic absorber
(polyurethane agglomerate) to attenuate acoustic vibrations. From
the outside the complete stabilization set-up is then a wooden
cube of 86 cm sides.

The stabilized temperature of the fiber spool has been studied. We
measure the temperature inside the box containing the fiber spool,
the floor temperature and the room temperature. The experimental
setup is located in an air conditioned laboratory at
22$\pm$1$^{o}$C.

\begin{figure}
  \includegraphics[width=\linewidth]{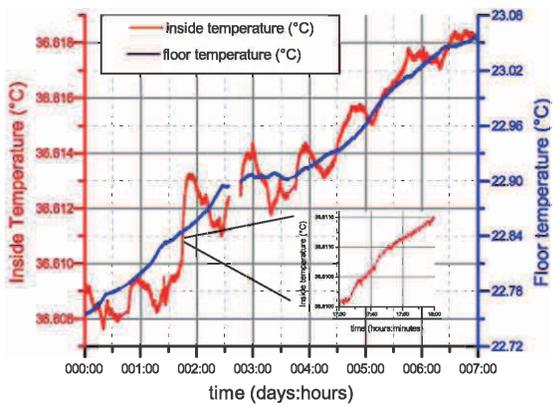}\\
  \caption{Temperature over 1 week for the box containing the fiber spools (red) and the floor (blue).}\label{fig_inside-floor}
\end{figure}

Figure ~\ref{fig_inside-floor} shows that, as expected, the
temperature of the box of the fiber spool is strongly correlated
on the long term with the temperature of the floor, that is used
for dissipating the heat inside the stabilization boxes. The
temperature drift for the floor temperature is 42.1~mK/day, while
for the inner box is 1.5~mK/day: the protection system attenuates
the long term temperature variations by a factor 27.

\begin{figure}
  \includegraphics[width=0.5\linewidth]{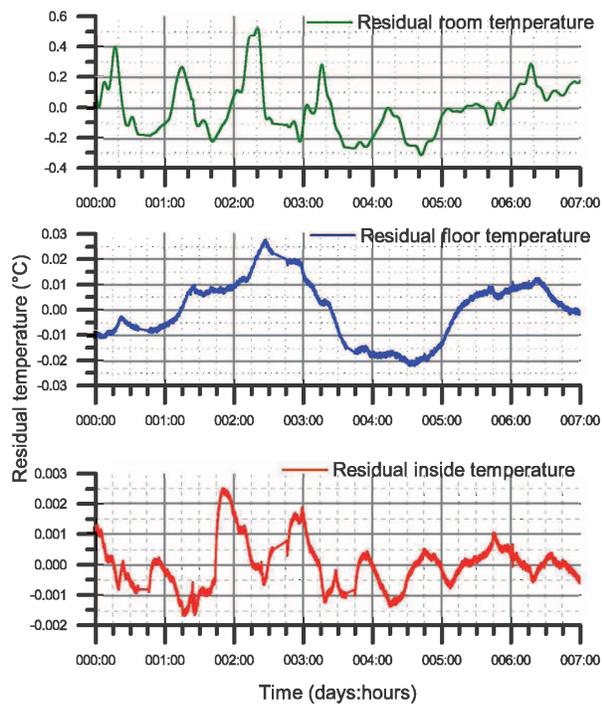}\\
  \caption{Residual fluctuations (after linear fit) of the smoothed room temperature (top, olive), floor temperature (middle, blue) and inner temperature (bottom, red).}\label{fig_residual_temp}
\end{figure}

However, the diurnal variations of the environment temperature
also affect the temperature of the fiber spool. By analyzing the
residual of the temperatures after removal of the linear drift
(fig.\ref{fig_residual_temp}), we can see that the room and box
temperatures have correlated diurnal variations, with the box
being delayed by about 15 hours, and attenuated by about a factor
200. Practically no diurnal variations are visible on the floor
temperature residuals.

We want to verify that the linear drift of the temperature of the
inner box corresponds to the linear frequency drift of the
stabilized Nd:YAG laser (a few hundred Hz/s) when measured with
respect to an independent outside reference (see section
\ref{drift_femto}). The relation between a temperature change
($\Delta\theta$) and a frequency change due to an optical length
variation of the fibre is
\begin{equation}\label{eq_drift2}
   \frac{\Delta \nu_{laser}}{\nu_{laser}} = \frac{2\Delta l}{l}=2S\Delta\theta
\end{equation}
where $S$ is the coefficient of optical length expansion of our
fibres $ S \sim 8 \times 10^{-6}$~K$^{-1}$ \cite{Dangui}. The
measured long term temperature drift over several days is
$\sim$~1.7~$\times 10^{-8}$~K/s (see above) giving a frequency
drift of only $\sim$80~Hz/s, somewhat less than observed. However,
the frequency measurements were carried out over much shorter
intervals, typically a few minutes. As the inset of figure
\ref{fig_inside-floor} shows, the temperature drift over such
short periods can reach up to $8 \times 10^{-7}$~K/s which is
compatible with the observed frequency drifts of a few hundred
Hz/s.
\section{Frequency stability of the lasers}
\label{stability}

The lasers are characterized in terms of power spectral density
(PSD) and Allan deviation of the fractional frequency fluctuations
of the lasers. To characterize the experimental stability of the
lasers, 2 different measurements have been performed.

The first one is the comparison between the 2 lasers of the
experiment, the local oscillator and the \mlaser{}. We aquire the
beat-note between the 2 stabilized lasers with a fast frequency
counter (Brilliant Instruments BI200). The counter can measure
from DC up to 2~GHz. The counter noise of the instrument depends
on the measured frequency, but is negligible in all our
measurements for Fourier frequencies below 100 kHz. The 2 lasers
are combined in a fiber coupler, and the output is focused on a
free space fast photo-diode. All the RF signals, and the frequency
counter have the same reference, so its frequency noise plays no
role in the measurement. Three measurements at different sampling
times are performed: at 1 $\mu$s, at 10 $\mu$s and at 1 ms. The
linear drift between the 2 stabilized lasers is of the order of a
few Hz/s over the measurement intervals (typically a few minutes).

The noise of the 2 stabilized lasers is correlated because the
stabilization system (spools, AOMs, diodes, etc..) are located in
the same box. As a consequence the absolute phase noise of the
lasers is certainly higher than when measured relative to each
other.

Therefore, the stabilized \mlaser{} has also been compared with a
completely independent system (a cryogenic sapphire oscillator at
12 GHz) using a frequency comb. The laboratories where the lasers
of the experiment and the frequency comb are located, are in
different buildings and are connected by a 130 m long optical
fiber link. The noise of the return optical fiber link, 260 m in
total, shows an Allan deviation of 1x10$^{-14}$ at 1 s and
5x$10^{-15}$ at 10 s, ie. is negligible for our purposes, at least
for integration times greater than $\sim$ 0.1 s.

The sampling time of the frequency comb measurement is 10 ms. The
linear drift of the \mlaser{} with respect to the frequency comb
is 600 Hz/s\label{drift_femto}. Compared to the drift between the
2 stabilized lasers, the drift of the frequency of the \mlaser{}
is about two orders of magnitude worse. This clearly shows that on
the long term our two lasers are strongly correlated, as expected.

\begin{figure}[!htbp]
  \includegraphics[width=1\textwidth]{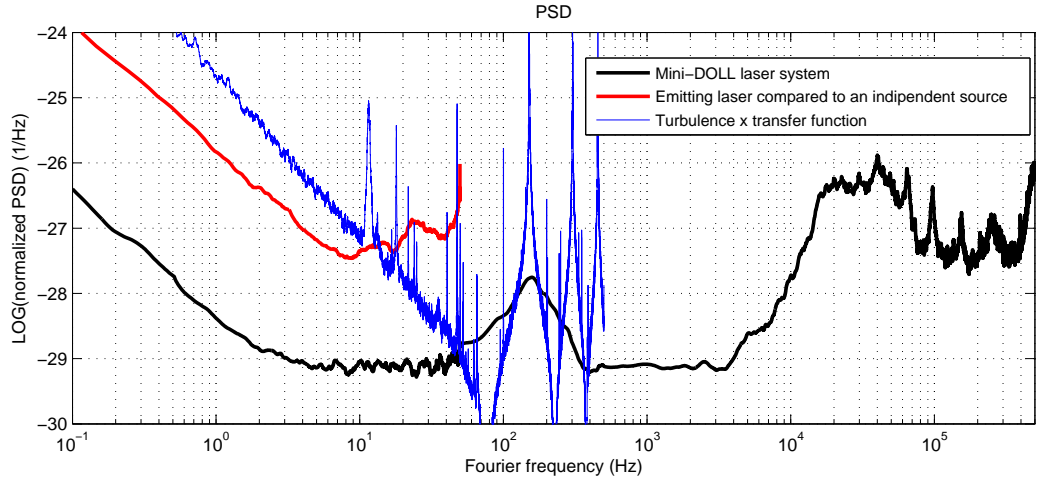}\\
  \caption{PSD of fractional frequency ($\Delta \nu/\nu_0$, $\nu_0$ = laser frequency = 282 THz). Black: beatnote between the 2 stabilized lasers. Red: beatnote between the stabilized \mlaser{} and an independent reference. Blue: expected noise of the turbulence weighted by the transfer function according to (\ref{eq_3}).}\label{fig_no_sweep_psd}
\end{figure}

\begin{figure}[!htbp]
  \includegraphics[width=\textwidth]{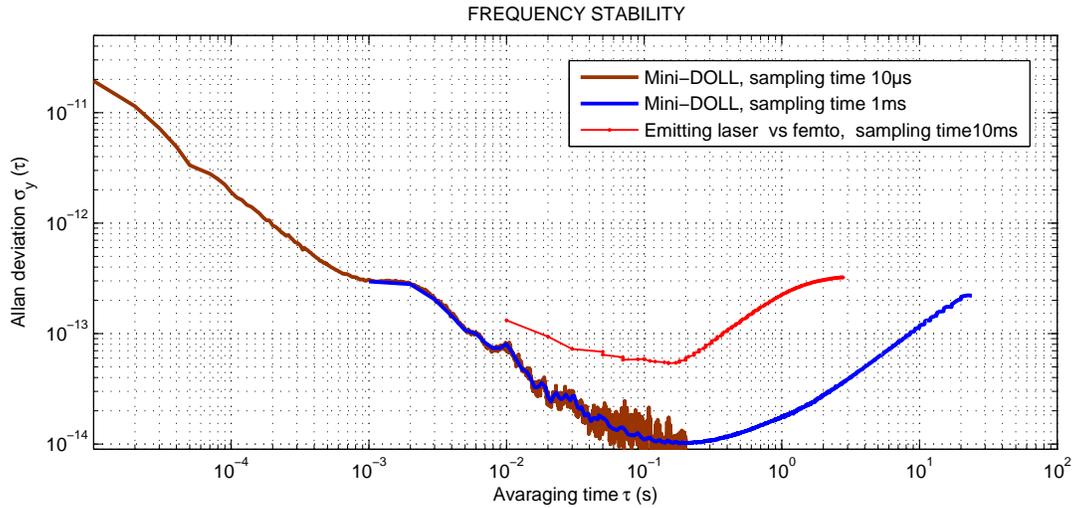}\\
  \caption{Allan deviation of the fractional frequency fluctuations between the two stabilized lasers and between the stabilized \mlaser{} and an independent reference.}\label{fig_no_sweep_allan_deviation}
\end{figure}

The performances of the stabilized lasers are shown in
fig.~\ref{fig_no_sweep_psd} and
\ref{fig_no_sweep_allan_deviation}. From figure
\ref{fig_no_sweep_allan_deviation} we see that for optimum
integration times ($\sim$ 0.2 s) the level of the noise of the
stabilized laser is $\sim 5$ times worse when compared to an
independent source than when the two lasers are compared between
them. For comparison we have also plotted in figure
~\ref{fig_no_sweep_psd} the expected noise that will limit our
satellite experiment, ie. the expected atmospheric turbulence
noise multiplied by the transfer function from equ. (\ref{eq_3})
($S_{atm}(f)(1+\cos (2\pi f T_{13}))/(1-\cos (2\pi f T_{13}))$
with $T_{13} = 6.66$ ms). We note that for Fourier frequencies of
interest to us ($<$10 Hz) the noise of the lasers will be
sufficient for our experiment.

The full experimental setup has been built and tested at the
SYRTE, Paris Observatory. In November 2012 it was transported to
the Calern site of the Observatoire de la c\^{o}te d'Azur in
southern France at an altitude of 1270 m, in preparation for the
satellite tests. In both sites the stabilized lasers remain locked
for some hours, but at the Calern site the locking is much more
robust maintaining lock for many hours without requiring any
adjustments. We have also noted that the typical stability of our
two lasers when compared to each other is in general about a
factor 2 to 3 better at Calern than in the Paris lab, although the
temperature stabilization of the room in Calern (just below the
telescope with open cupola etc..) is significantly less good than
in the Paris laboratory. Both these effects are most likely due to
the lower atmospheric pressure at the Calern altitude leading to
less thermal and acoustic coupling between the room and the inside
box that contains the fiber spools for the stabilization.
\section{Linear frequency sweep}
\label{linear}

Next we have verified that the laser stability is not modified
when the local oscillator laser frequency is swept over 20 GHz at
different rates. As a first step we carry out purely linear
frequency sweeps, which also serve for the precise determination
of the delay $\tau$ (c.f. equ. (\ref{eq_302})) required for the
controlled non-linear sweep for correct Doppler compensation.

To perform the following measurements, the beat-note between the
swept local oscillator and the \mlaser{} is collected with a 25
GHz fast photodiode. The signal from the photo-detector is divided
by 8 by a low noise digital divider that operates from some kHz to
18 GHz input frequency. The divided signal is then recorded with
the frequency counter described in the previous section, that
operates up to 2 GHz. The setup allows to measure a frequency
change of 32GHz, from -16 GHz to +16 GHz.

\begin{figure}
  \includegraphics[width=0.5\linewidth]{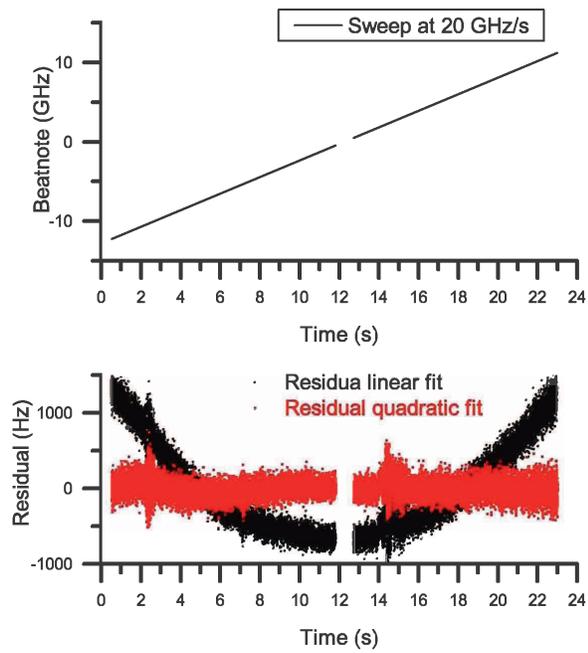}\\
  \caption{Top: beatnote ($\nu_{emit}-\nu_{LO}(t)$) between the 2 lasers when the local oscillator (LO) is linearly swept at 1 GHz/s over more than 20 GHz. Bottom: residuals of the beatnote after linear (black) and quadratic (red) fitting.}\label{figure_tau}
\end{figure}

Figure~\ref{figure_tau} shows a linear frequency sweep at 1 GHz
spanning more than 20 GHz. The residuals clearly show a quadratic
behavior, as expected from the dispersion of the fibre (for our
Corning HI1060 fibres the nominal dispersion coefficient around
1060 nm is 1.43$\times 10^{-22}$ s$^2$/km). We fit a quadratic
model of the local oscillator (LO) frequency to the data, of the
form

\begin{equation}\label{QuadFit}
\nu_{LO}(t) = \nu_0+Ct+Dt^2
\end{equation}

obtaining C=-1043714018.62$\pm$0.6 Hz/s and D=-15.470$\pm$0.02
Hz$^2$/s. Combining equations (\ref{eq_302}) and (\ref{QuadFit})
we find the relation for the dispersion effect on our delay
$d\tau/d\nu = -2D\tau_0/C^2$ where $\tau_0$ is the delay of our 2
km (1 km double pass) fibre at the nominal frequency\footnote
{Substituting the derivative of (\ref{QuadFit}) into
(\ref{eq_302}) with $\tau(\nu)=\tau_0+(d\tau/d\nu)(\nu-\nu_0)$ we
obtain $C\tau_0+(2D\tau_0+C^2(d\tau/d\nu))t+{\cal O}(t^2,t^3)=
\Delta\nu_{rf}$. Knowing that $\Delta\nu_{rf}$ is constant for all
$t$ we have $2D\tau_0+C^2(d\tau/d\nu) = 0$. }. We find a
dispersion coefficient of 1.36$\times 10^{-22}$ s$^2$/km, very
close to the value quoted by the manufacturer.

\begin{figure}[!htbp]
  \includegraphics[width=0.5\linewidth]{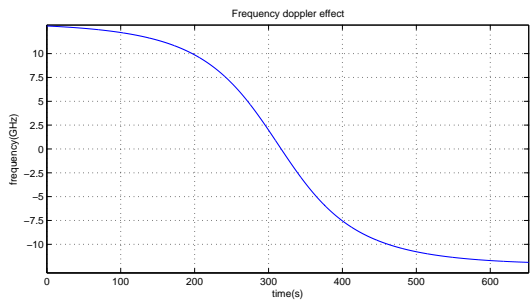}\\
  \caption{Simulated Doppler frequency shift for a typical satellite pass}\label{fig_simulation_envisat}
\end{figure}

In principle it is possible to take the dispersion effect into
account when programming $\Delta\nu_{rf}(t)$ to compensate for the
predicted Doppler effect of the satellite. However, in practice
this is not necessary as the sweep velocities are low when the
frequency is far (10 GHz) from its nominal value (see section \ref
{non-linear} and fig. \ref{fig_simulation_envisat}). Consequently
an error in $\tau$ has a smaller effect at those frequencies ie.
when the dispersion effect is largest. For example, at a 10 MHz/s
sweep rate a 3 ps error in tau ($=d\tau/d\nu \times 10$ GHz)
corresponds to a 3 Hz/s sweep error, less than the natural drift
of our lasers.

Thus we determine a single value of $\tau$ at the central
frequency and use it to program $\Delta\nu_{rf}$ for the satellite
Doppler correction. To do so the following procedure has been
adopted. In the stabilization system of fig.~\ref{fig_rf},
constant values of $\Delta\nu_{rf}$ are introduced in the
servo-loop. This corresponds to linearly sweeping the local
oscillator. To take into account the natural drift of the laser,
for each value of $\Delta\nu_{rf}$, we sweep with the positive
value and the negative value. To further take in account the
natural drift of the laser, we repeat the same values of
$\nu_{rf}$ after some seconds. The set of data of $\Delta\nu_{rf}$
is the following: +1kHz, -1 kHz, then after 1 second +2 kHz, -2
kHz, +5 kHz and -5 kHz. After 50 seconds we repeat +2 kHz and -2
kHz, and after waiting 50 seconds we apply +5 kHz and -5 kHz. The
values of $\Delta\nu_{rf}$ sent to the system, 1 kHz, 2 kHz and 5
kHz, correspond to laser frequency changes of roughly 100, 200 and
500 MHz/s, respectively.

For each $\Delta\nu_{rf}$ value, the laser beatnote is recorded
and a linear fit of the data is carried out to compute the values
of $\tau$ from which we calculate the average value and its
standard deviation. A typical averaged value of $\tau$ is
9.5761890718608$\times 10^{-6}$ s with a standard deviation of
2$\times 10^{-12}$ s. We attribute that uncertainty to the effect
of the natural drift of the two lasers during and between
measurements. Repeating the measurements, even separated by
several days, the obtained value of $\tau$ does not change by more
than a few picoseconds.

\begin{figure}
  \includegraphics[width=\linewidth]{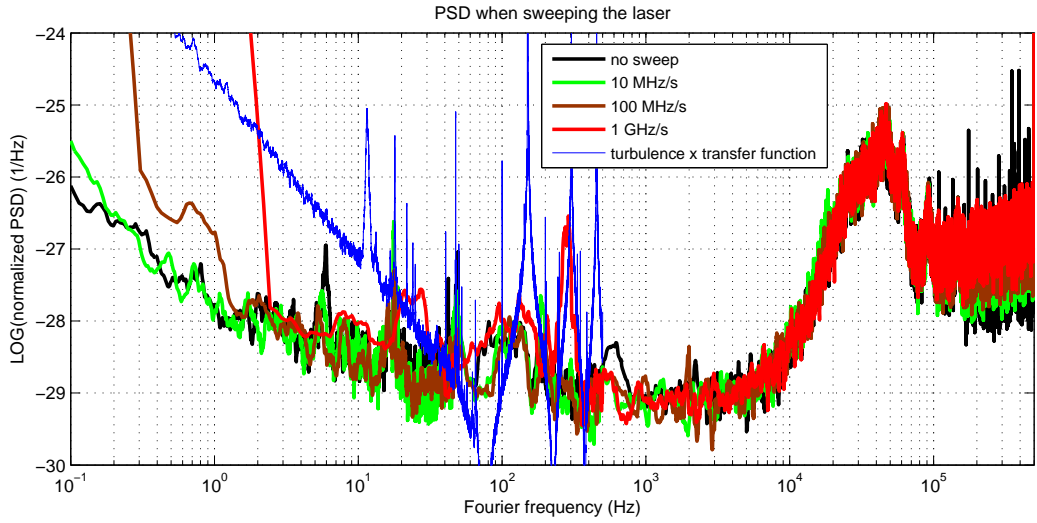}\\
  \caption{Fractional frequency PSD of the beatnote between the 2 lasers when the local oscillator is linearly swept up to 1 GHz/s, together with the expected noise of turbulence weighted by the transfer function according to (\ref{eq_3}).}\label{figure_sweep3}
\end{figure}

The instability at different linear sweep velocities has been
tested. Fig.~\ref{figure_sweep3} shows the PSD of the fractional
frequency difference between the 2 lasers, after subtraction of
the applied linear sweep. The figure shows that it is possible to
linearly sweep the laser up to 1 GHz/s without significantly
degrading the stability. The data comply with the requirement for
measuring the turbulence, even taking in account the worsening of
the noise level when comparing to an independent laser source (see
fig.~\ref{fig_no_sweep_psd}).

Note that the data for fig.~\ref{figure_sweep3} were collected
months before the data of fig.~\ref{fig_no_sweep_psd}. The new
data have been obtained after several improvements of the
isolation of the boxes that contain the fiber spools, which
resulted in the disappearance of the peaks in the PSD below 20Hz,
probably due to pressure fluctuations of the room.

In summary, we have carried out linear sweeps up to 1 GHz/s over
more than 25 GHz without a noticeable degradation of the laser
stability and in compliance with our requirements for the
satellite experiment ie. with a noise level below the expected
contribution from atmospheric turbulence.
\section{Non-linear frequency sweep}
\label{non-linear}

To evaluate the capability of our experimental setup to compensate
the satellite Doppler effect, we have simulated the corresponding
frequency shift for a typical satellite pass for our experiment
(800 km altitude, inclination 98 degrees, pass culminating at 78
degrees elevation), shown in fig.~\ref{fig_simulation_envisat}.
The maximum sweep rate is $\sim$120 MHz/s, and it corresponds to a
decrease of the frequency of $\sim$24.8 GHz. The full duration is
652~seconds.

We measure the beat-note between the two stabilized lasers as in
the previous section. After stabilizing the 2 lasers, the
beat-note is first increased by +12GHz by use of the DDS acting on
the PZT of the local oscillator. Next we sweep the local
oscillator frequency by introducing the correct data sequence in
the DDS. The sampling time for the counter is 1 ms.

As a result, the frequency measured by the counter decreases, then
it crosses zero and it restarts to increase up to to $\sim$12 GHz.
This is due to the fact that the frequency counter measures the
absolute value of the frequency difference between the 2 lasers,
regardless of the sign. In the measuring setup, there are several
components that cut the low frequency resulting in the absence of
data from the counter near 0 frequency.

\begin{figure}
  \includegraphics[width=0.5\linewidth]{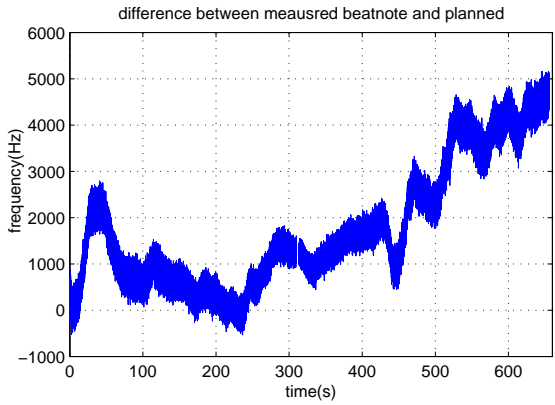}\\
  \caption{Difference between the measured beatnote and the frequency Doppler shift simulation, $\nu_{diff}$.}\label{fig_difference_simulation}
\end{figure}

The data from the counter, $\nu_{measured}$ are processed in the
following way. We compute the difference
$\nu_{diff}=\nu_{measured}-\nu_{prevision}$ after interpolating
the data to ensure that $\nu_{measured}$ and $\nu_{prevision}$ are
taken at the same instant (see
fig.\ref{fig_difference_simulation}). The frequency difference
$\nu_{diff}$ represents the error we commit in our Doppler
compensation and it includes the natural drift of the beatnote.
The remaining linear drift is 5.76~Hz/s.

\begin{figure}
  \includegraphics[width=0.5\linewidth]{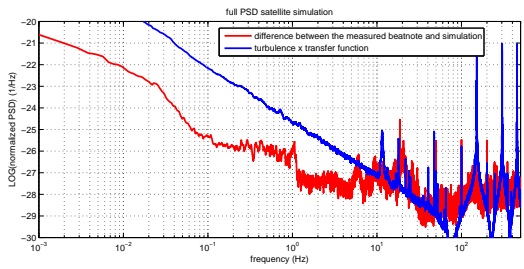}\\
  \caption{Fractional frequency PSD of the difference between the measured beatnote and the Doppler shift simulation, $\nu_{diff}$ (red), together with the expected noise of turbulence weighted by the transfer function according to (\ref{eq_3}) (blue).}\label{fig_simulationPSD}
\end{figure}

\begin{figure}
  \includegraphics[width=\linewidth]{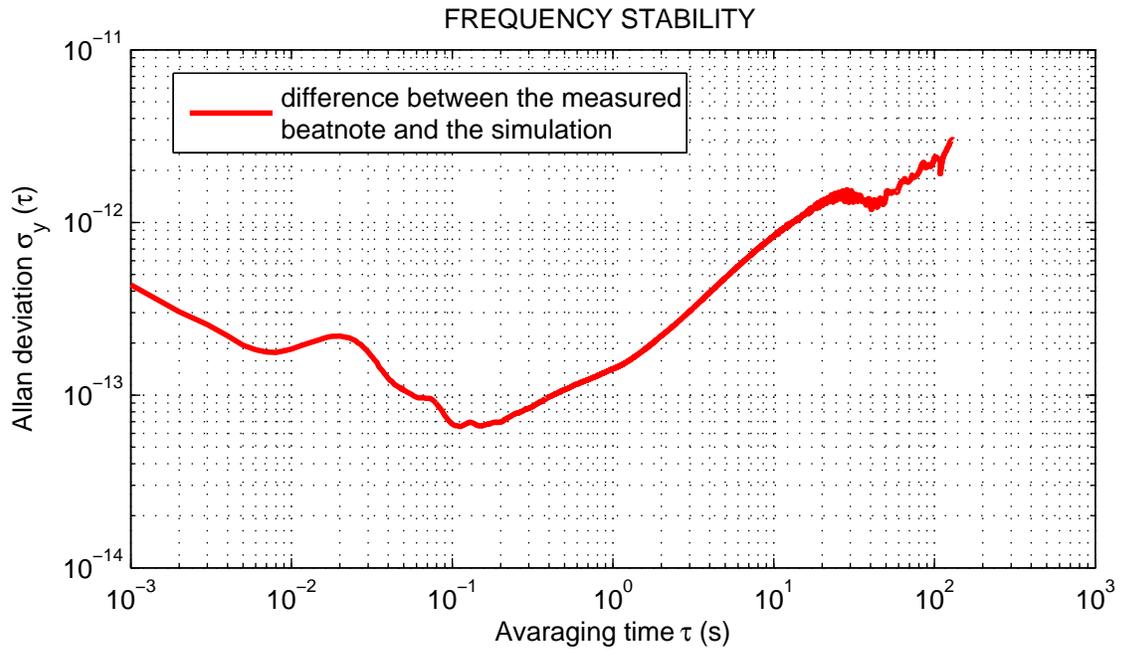}\\
  \caption{Allan deviation of the difference between the measured beatnote and the Doppler shift simulation, $\nu_{diff}$.}\label{fig_simulationALLAN}
\end{figure}

In fig. \ref{fig_difference_simulation}, we see the natural
relative drift of our lasers but the large signature of the
Doppler effect is entirely compensated, ie. the actual frequency
sweep corresponds to the intended one to within the natural drift
of our lasers. This result also justifies neglecting the
dispersion effect (see section \ref{linear}) for our purposes.
When computing the PSD and the Allan Deviation of $\nu_{diff}$
(Fig.\ref{fig_simulationPSD} and \ref{fig_simulationALLAN}) we see
slightly larger noise than when applying a linear sweep or no
sweep, but which is still sufficiently low for our purposes.
\section{Frequency stability of the amplifier}
\label{amplifier}
To maximize the returning signal power we amplify the signal from
our stabilized \mlaser{} using a multiple stage Nufern fibre
amplifier (max. 50 W output). Typically such amplifiers are
specified for a minimum seed laser linewidth of about 10 kHz, but
our stabilized laser has much lower linewidth (around 25 Hz) and
we require the amplified light to have the same stability
characteristics as the seed laser. To ensure that this is the case
we have tested a 10 W model of the same amplifier (on loan from
Nufern) before acquiring our 50 W version. As shown below, the
phase noise added by the amplifier is negligible for our purposes,
so we use it to amplify our emission laser as the last stage
before injection into the telescope. If this had not been the case
we could have stabilized the laser+amplifier (extracting a small
fraction of light before the telescope) on our fibre spool at the
cost of higher complexity. It turns out that this is not
necessary.

\begin{figure}[!htbp]
  \includegraphics[width=0.5\linewidth]{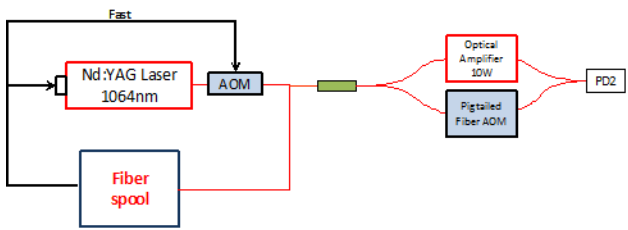}\\
  \caption{Experimental setup for measuring the noise of the optical amplifier.}\label{fig_setup_amplified}
\end{figure}

For the amplifier phase noise measurement we have used a
Mach-Zender setup as shown in figure \ref{fig_setup_amplified}.
The \mlaser, that is frequency stabilized, is split in 2 parts:
one part passes in a fiber pigtailed AOM, which shifts its
frequency by 80 MHz, the second is amplified and 1 mW is split off
the amplified beam to perform the measurement. The beat-note at 80
MHz is measured as described in sect. \ref{set-up}.

\begin{figure}
  \includegraphics[width=0.5\linewidth]{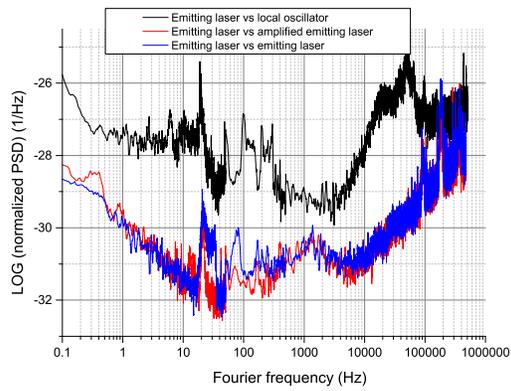}\\
  \caption{Fractional frequency PSD. Black: \mlaser{} compared to local oscillator (for reference). Red: with amplifier. Blue: Without
amplifier ("short circuiting" the amplifier in figure 3), the blue
line corresponds to the noise floor of our measurement
system.}\label{fig_results_amplifier}
\end{figure}

Fig. \ref{fig_results_amplifier} shows the measurements done with
and without amplifier (short-circuiting the amplifier in fig.
\ref{fig_setup_amplified}), the latter characterizes the noise
floor of our measurements system.

From fig. \ref{fig_results_amplifier} we conclude that the PSD of
the frequency noise added by the amplifier is at or below
10$^{-31}$ Hz$^{-1}$ in fractional frequency ($<$ 8 x 10$^{-3}$
Hz$^{2}$/Hz in absolute frequency) for 10 Hz $< f <$ 10 kHz, and
well below the noise of our stabilized \mlaser{} at all
frequencies. The peak at 20 Hz is most likely due to pressure
fluctuations in the laboratory (which we have investigated
seperately) and the noise above $\sim$ 10 kHz is consistent with
the phase noise of our frequency counter. It seems clear that the
observed noise is the noise floor of the measuring setup, so the
actual phase noise added by the amplifier is likely to be below
that. For comparison, a flat PSD (white frequency noise) at 8 x
10$^{-3}$ Hz$^{2}$/Hz corresponds to a line-width of 25 mHz, ie.
we expect that the amplifier does not degrade any seed signal as
narrow as that or below, which is remarkable.
\section{Conclusion}
We have presented an original laser system designed for ground to
low Earth orbit satellite coherent optical links. The aim was to
develop a system that is capable to cope with the high Doppler
dynamics ($\pm$12 GHz with rates up to 120 MHz/s) whilst
displaying state of the art frequency stability in order to be
limited by phase noise from atmospheric turbulence rather than the
laser stability. Our solution is based on laser stabilization
using unequal arm fibre interferometers closely inspired by
\cite{Jiang}. Based on simulated satellite Doppler shifts and
expected turbulence conditions taken from \cite{Djerroud} we
demonstrate that our system satisfies these requirements. A system
with better stability was reported in \cite{Jiang}, however with
sweep rates no higher than 40 MHz/s and range $< 1$ GHz, that are
insufficient for our purposes. Much broader and faster tuning was
reported in \cite{Roos}, however with insufficient stability for
our purposes. We combine both these features in a compromise
between performance and compactness as it has to be moved
frequently to astronomical sites where it will be used to realize
first coherent ground-satellite optical Doppler ranging.

The system has been transported from Paris to the Observatoire de
la c\^{o}te d'Azur in southern France where it was used with the
1.5 m lunar/satellite ranging telescope. First tests have
encountered difficulties because of the impossibility of using the
in situ adaptive optics system as planned. Further tests are
planned for 2014 after an upgrade of the adaptive optics system.

\section{Acknowledgements} Helpful discussions with Giorgio Santarelli and S\'{e}bastien
Bize are greatfully acknowledged. We would also like to thank Yann
Le Coq, Laurent Volodimer, Jos\'{e} Pinto and Michel Lours for
their valuable contributions. This work was supported by CNES
research grant (R-S09/SU-0001-021) DA 10069354, financing by LNE
and by the Action Sp\'{e}cifique GRAM (INSU/INP/CNES).

\newpage
%
%
%
%
%
%
%
%
%
%
%
%
\end{document}